\documentclass[letterpaper,titlepage,11pt]{article}
\pdfoutput=1
\usepackage{hyperref}
\usepackage{epsfig}
\usepackage{graphicx}
\usepackage{amsmath}
\usepackage{amsfonts}
\usepackage{amssymb}
\usepackage{mathrsfs}
\usepackage{fancybox}
\usepackage{ragged2e}
\usepackage{tikz}
\usepackage{wasysym}
\newcommand{\mycite}[1]{~{\cite{#1}}}
\newcommand{\myfigref}[1]{~{Fig.~(\ref{#1})}}
\newcommand{\myref}[1]{~{(\ref{#1})}}

\newcommand{\beq}{\begin{equation}}
\newcommand{\eeq}{\end{equation}}
\newcommand{\bw}[1]{\protect{\bf #1}}

\begin{document}

\begin{titlepage}
\begin{flushright}
ITEP-TH-19/12
\end{flushright}

\rightline{\vbox{\small\hbox{\tt } }}
 \vskip 1.8 cm

\centerline{\LARGE \bf Anomalous Zero Sound} \vskip 1.5cm

\centerline{\large {\bf A.Gorsky$\,^{1}$},  {\bf
A.~V.~Zayakin$\,^{2,1}$}}

\vskip 1.0cm

\begin{center}
\sl $^1$ Institute of Theoretical and Experimental Physics,\\
B.~Cheremushkinskaya ul. 25, 117259 Moscow, Russia\\
\vskip 0.4cm
\sl $^2$ Dipartimento di Fisica, Universit\`a di Perugia,\\
I.N.F.N. Sezione di Perugia,\\
Via Pascoli, I-06123 Perugia, Italy
\end{center}
\vskip 0.6cm

\centerline{\small\tt a.zayakin@gmail.com }

\vskip 1.3cm \centerline{\bf Abstract} \vskip 0.2cm \noindent We
show that the anomalous term in the current, recently suggested by
Son and Yamamoto, modifies the structure of the zero sound mode
in the Fermi liquid in a magnetic field.
\end{titlepage}

\addtocontents{toc}{\protect\setcounter{tocdepth}{1}}

\setcounter{page}{1}
\section{Waves in Fermi Liquids}
The standard theory of a Fermi liquid\mycite{Landau1957} allows us
to predict a lot of interesting collective phenomena in quantum
liquids, in particular, the existence of different modes of
density fluctuations.

The main tool of the analysis of Fermi
liquids is the kinetic equation
\beq\label{kinetic} \frac{\partial n(\bw{p},\bw{x})}{\partial
t}+\frac{\partial \epsilon(\bw{p},\bw{x})}{\partial \bw{p}
}\frac{\partial n(\bw{p},\bw{x})}{\partial \bw{x}}-\frac{\partial
\epsilon(\bw{p},\bw{x})}{\partial \bw{x} }\frac{\partial
n(\bw{p},\bw{x})}{\partial \bw{p}}=0 \eeq
Here and below $n(\bw{p},\bw{x})$ is the density of particles in
the phase space, and $\epsilon(\bw{p},\bw{x})$ is the energy
density. It is important to understand that the energy density in
a specified mode is not equivalent to the dispersion law for the
free mode, since we deal with interacting systems, wherein the
whole system ``contributes energy'' into the given mode with
momentum $\bw{p}$ and position $\bw{x}$. The other crucial thing
for understanding Fermi liquids is that the natural basis for the
Hilbert space of excited states is the quasiparticle basis. The
quasiparticle vacuum is the state with all energy levels below the
Fermi surface are filled with particles. Thus while we deal with
small excitations, the quasiparticles have the energy
$\epsilon=\epsilon_F$ and momentum $\bw{p}$ confined to the Fermi
sphere.

The kinetic equation can be understood either as the semiclassical
limit of the evolution equation for the density operator $n$ by
means of the Hamiltonian $H$
\beq \frac{\partial n}{\partial t}=\frac{i}{\hbar}[n,H], \eeq
with the classical Poisson bracket of $\{\epsilon,n\}$ emerging
from the commutator $[n,H]$, or as the classical continuity
equation for the current
\beq \frac{\partial n(\bw{x})}{\partial t}+\frac{\partial
\bw{j}(\bw{x})}{\partial \bw{x}}=0,\eeq where the current is
normally given by
\beq\label{current} \bw{j}(\bw{x})=-\int \frac{d^3p}{(2\pi)^3}
\epsilon({\bw{p},\bw{x}})\frac{\partial n(\bw{p},\bw{x})
}{\partial p}=\int \frac{d^3p}{(2\pi)^3}
n(\bw{p},\bw{x})\frac{\partial \epsilon(\bw{p},\bw{x}) }{\partial
p}=n(\bw{x})\bw{v},\eeq
where we dropped the boundary terms under the integral sign.

The basic assumption for the standard Fermi liquid theory is that
the energy in the $\bw{p}$ mode ``feels'' the fluctuations of the
rest of the liquid. Thus we can say that this model starts from a
self-consistent density distribution of interacting
quasiparticles, the interaction strength not necessarily being
small. While the ``equilibrium'' energy density
$\epsilon_0(\bw{p})$ remains function of momentum only, the energy
fluctuation $\epsilon(\bw{p},\bw{x})$ is a function of both
momenta and coordinates
\beq \epsilon(\bw{p},\bw{x})=\epsilon_0(\bw{p})+\delta
\epsilon(\bw{p},\bw{x}), \eeq
and is related to the density fluctuations $\delta n$, which, in
turn, are defined as
\beq n(\bw{p},\bw{x})=n_0(\bw{p})+\delta n(\bw{p},\bw{x}), \eeq
by means of a convolution with the ``interaction kernel''
$f(\bw{p},\bw{p}')$
\beq \delta \epsilon(\bw{p},\bw{x})=\int
\frac{d^3p'}{(2\pi\hbar)^3} f(\bw{p},\bw{p}')\delta
n(\bw{p},\bw{x}), \eeq
which, assuming the quasiparticle density localized on the Fermi
sphere, becomes
\beq \delta \epsilon(\bw{p},\bw{x})=\int d^2\Omega\,
F(\theta,\theta') \delta n(\theta',\bw{x}), \eeq
where the dimensionless formfactor function $F(\theta,\theta')$
($F=\frac{p_F m}{2\pi^2 \hbar^3}f$) can be represented as a series
in the spherical harmonics. Actually the interaction kernel $f$ is
a matrix that includes spin-spin interaction formfactor $\sim \bf
\sigma\sigma'$; the integral sign in that case must be understood
as a trace over the spin indices as well. We omit the spin indices
from our analysis, since we limit ourselves to the class of
interactions with $f\sim \hat{1}$;
The equilibrium density and energy distributions $n_0$,
$\epsilon_0$ are time- and coordinate-independent. If $\delta n$
describes a space-dependent fluctuation of particle density, the
kinetic equation\myref{kinetic} will become
 \beq\label{kinetic1}\frac{\partial \delta n}{\partial
t}+\frac{\partial \epsilon_0}{\partial \bw{p} }\frac{\partial
\delta n}{\partial \bw{x}}-\frac{\partial \delta \epsilon
}{\partial \bw{x} }\frac{\partial n_0 }{\partial \bw{p}}=0. \eeq
Now, consider a flat-wave fluctuation of the particle density
\beq \delta n = \delta(\epsilon-\epsilon_F) \nu(\theta,\phi)
e^{i(\omega t - \bw{k}\bw{r})}. \eeq
Taking into account that
\beq \frac{\partial n_0(\bw{p},\bw{x})}{\partial
\bw{p}}=-\bw{v}_F\delta(\epsilon-\epsilon_F),\eeq
we get the following integral equation from\myref{kinetic1}
\beq\label{e12} (\omega -
\bw{k}\bw{v})\nu(\theta)=\bw{k}\bw{v}\int \frac{d^2\Omega'}{4\pi}
F(\theta,\theta',\phi,\phi')\nu(\theta'), \eeq
where $d^2\Omega=\sin\theta d\theta d\phi$. Restricting our
analysis to $\phi$-independent modes solely, choosing $\bw{k}$ as
the $Oz$ direction on the Fermi sphere, noticing that
$\bw{k}\bw{v}=kv_F\cos\theta$, introducing a convenient parameter
$s\equiv \frac{\omega}{k v_F}$, we get
\beq\label{zerosound}
\left(s-\cos\theta\right)\nu(\theta)=\cos\theta\int
\frac{\sin\theta' d\theta'}{2}\nu(\theta')F(\theta,\theta').  \eeq
Let us limit ourselves with the the zeroth and the first harmonics
of $F$
\beq\label{kinetic2} F=F_0+F_1\left(\cos\theta\cos\theta' +
\sin\theta\sin\theta' \cos(\phi-\phi')\right). \eeq
This will mean that a solution to\myref{kinetic2} must be sought
in the form
\beq \nu(\theta)=(C_0+C_1
\cos\theta)\frac{\cos\theta}{s-\cos\theta}, \eeq
where $C_0,C_1$ are some constants which are subject to conditions
\beq
\begin{array}{l}
\displaystyle C_0= F_0 \int \frac{\sin\theta d\theta}{2}(C_0+C_1
\cos\theta)\frac{\cos\theta}{s-\cos\theta}, \\ \\
\displaystyle C_1= F_1 \int \frac{\sin\theta
d\theta}{2}\cos\theta(C_0+C_1
\cos\theta)\frac{\cos\theta}{s-\cos\theta}.
\end{array}
\eeq
This system of linear equations upon $C_0,C_1$ must have a zero
determinant to be solvable, which imposes a condition upon $s$
thus providing us with a dispersion relation following from the
equation
\beq\label{det} \left(\frac{1}{2} s \log
\left(\frac{s+1}{s-1}\right)-1\right) \left(F_0 (F_1+3)+3 F_1
s^2\right)=F_1+3.\eeq
The case of small $F_0,F_1$ (both are dimensionless formfactors,
describing interaction of quasiparticles; they can be directly
inferred e.g. from the four-particle interaction Hamiltonian) may
be called weakly interacting phase, whereas the case of large
formfactors -- strongly interacting. In the weakly interacting
phase we expand the equation\myref{det} around $s=1$ and get the
dispersion law
\beq \label{disp1}s=1+2 e^{-\frac{2 (F_0 (F_1+3)+4
F_1+3)}{F_0(F_1+3)+3 F_1}}, \eeq
whereas for the strongly-interacting case the expansion takes
place around $s=\infty$ and we get
\beq
\label{disp2}s=\frac{1}{\sqrt{3}}\sqrt{F_0+\frac{3F_1}{5}+\frac{F_0
F_1}{3}}. \eeq
The oscillation mode we have described so far is known as the zero
sound mode. It is important for us that this phenomenon is equally
successfully realized both in the framework of the field theory
(Landau Fermi liquid theory, explicated above) and in
holography\mycite{Karch:2008fa}. Of importance is also the fact
that holography, advertised as the appropriate language for the
strongly-coupled theory, indeed reproduces the dispersion
law\myref{disp2} for the large $F_0,F_1$, rather than the
weakly-coupled\myref{disp1} regime. In particular, even the
quantum attenuation contribution (the absorptive part in the
dispersion law not shown here) is reproduced form holography as
well\mycite{Karch:2008fa}. Let us emphasize that zero sound is not
the single quasiparticle excitation but the collective excitation
of the Fermi surface.

Now we proceed to an interesting modification of zero sound due to
an anomaly in the magnetic field.

\section{Anomaly and Chirality}
Recently Son and Yamamoto\mycite{Son:2012wh} have suggested that
in theories with anomalies the expression for the
current\myref{current} be supplied with two extra terms
\beq \label{current} \bw{j}(\bw{x})=\int \frac{d^3p}{(2\pi)^3}
\left[-\epsilon(\bw{p},\bw{x})\frac{\partial n(\bw{p},\bw{x})
}{\partial p} - \left(\bw{\Omega}\frac{\partial
n(\bw{p},\bw{x})}{\partial \bw{p}}\right)\epsilon(\bw{p},\bw{x})
\bw{B} -\epsilon(\bw{p},\bw{x}) \left(\bw{\Omega}\times
\frac{\partial n(\bw{p},\bw{x})}{\partial \bw{x}}\right) \right].
\eeq
Here the crucial element of the construction is the ``dual''
magnetic field strength $\bw{\Omega}$ in the momentum space.

This extra current leads to the existence of several interesting
effects in the presence of an external magnetic field. One of them
is the chiral magnetic effect\mycite{Fukushima:2008xe}, which
amounts to generation of an electric current along a magnetic
field in presence of an axial chemical potential $\mu_A$
\beq \bw{j}=\frac{\mu_A \bw{B}}{2\pi^2}.\eeq
This phenomenon was realized in the holographic approach to QCD in
several different ways\mycite{Yee:2009vw, Rebhan:2009vc,
Gorsky:2010xu} and observed on lattice\mycite{Buividovich:2009wi}.
Holography models the chiral magnetic effect via a Chern-Simons
term
\beq \displaystyle S_{CS}=-\frac{N_c}{24\pi^2}\int d^5x
\epsilon^{\mu\nu\lambda\rho\sigma}A_\mu
F_{\nu\lambda}F_{\rho\sigma},\eeq
which generates the anomalous current.
Recently the same term in the holographic action has been shown to
be responsible for another very interesting effect, the chiral
magnetic wave\mycite{Kharzeev:2010gd}. The essence of this effect
is propagation of a chirality wave through a medium in a magnetic
field ${\bw{B}}=(0,0,B)$
\beq n_A\,\, \sim \,\, e^{i (\omega t - k z)}.  \eeq

The anomalous term in the current in the magnetic case has the
immediate counterpart for the rotating Fermi fluid. This is the
generic situation when the following substitution $B\rightarrow
2m\omega$ yields the correct anomaly for the rotating matter. In
particular the chiral magnetic effect has the parallel chiral
vortical effect. The presence of the proper term in the current in
the rotating case has been remarked in \cite{Kirilin:2012sd}. In
what follows we have in mind that the anomalous zero sound
splitting can be equally considered for the magnetic or rotating
cases.

\section{Anomalous Current}

The  modification of the current\myref{current} allows us to
include the anomaly-driven dynamics into the kinetic equation.
The realization of the
anomaly as a contribution to the current via Berry phase
in\mycite{Son:2012wh} has paved us the way to understanding how
the modes of density oscillations in Fermi liquids are modified
due to anomalies in a magnetic field. A geometrical argument for
the appearance of the Wess-Zumino term for a Fermi surface due to
a Berry phase was suggested recently by
Zahed\mycite{Zahed:2012yu}.
Consider a Fermi liquid with right and left quasiparticles,
described by their densities $n_R$ and $n_L$. We have now two
currents

\beq \begin{array}{l}\label{current} \displaystyle
\bw{j}_{R,L}(\bw{x})=\int \frac{d^3p}{(2\pi)^3}
\left[-\epsilon_{R,L}(\bw{p},\bw{x})\frac{\partial
n_{R,L}(\bw{p},\bw{x}) }{\partial p} -
\left(\bw{\Omega}_{R,L}\frac{\partial
n_{R,L}(\bw{p},\bw{x})}{\partial
\bw{p}}\right)\epsilon_{R,L}(\bw{p},\bw{x}) \bw{B}-\right.\\
\left. \displaystyle
\hphantom{\bw{j}_{R,L}(\bw{x})=}-\epsilon_{R,L}(\bw{p},\bw{x})
\left(\bw{\Omega}_{R,L}\times \frac{\partial
n_{R,L}(\bw{p},\bw{x})}{\partial \bw{x}}\right) \right],
\end{array}\eeq
The oscillations of the left and right densities, that to be
identical and free one from each other become now coupled via the
anomaly term. This will lift the degeneracy in the dispersion law,
yielding two different types of zero sound. This phenomenon will
be called by us ``anomalous zero sound''.

One might question the validity of such an approach given that we
combine the non-relativistic physics of quasiparticles living on a
Fermi surface, and an explicitly relativistic extra current by
Son-Yamamoto. Indeed, since the extra current terms are derived
from the Berry phase, they require the presence of a monopole at
the origin of the momentum space. Son and Yamamoto have clearly
related the extra term in the current with the anomaly. Since the
anomalous contribution exists both for massive and massless
particles we should keep the monopole at the origin of the
momentum space for the massive particles as well as Son and
Yamamoto did.

Alternatively, one could appeal to the relativistic formulation of
the Fermi liquid theory by Baym and Chin\mycite{Baym:1975va},
which has been shown to be much similar to the original Landau
non-relativistic formalism. Then the argumentation by Son-Yamamoto
based on Berry's phase would be directly applicable.

The ``dual'' magnetic field $\bw{\Omega}$ is given by the
condition that its flux through the Fermi surface is the
topological charge
\beq \frac{1}{2\pi}\int d\bw{S}\,\, \bw{\Omega}= \pm 1, \eeq
whence we can choose for $\bw{\Omega}$ a hedgehog Ansatz
\beq \bw{\Omega}_{R,L}=\pm \frac{\bw{n} }{2p_F^2}, \eeq
where $p_F$ is the Fermi momentum, $\bw{n}$ is the unit normal
vector on the Fermi surface, $ d\bw{S}$ is the integration measure
on the Fermi surface.

The extra contributions in the kinetic equation $\frac{\partial
n(\bw{x})}{\partial t}+\frac{\partial \bw{j}(\bw{x})}{\partial
\bw{x}}=0$ that come from $\left(\bw{\Omega}\frac{\partial
n(\bw{p},\bw{x})}{\partial \bw{p}}\right)\epsilon(\bw{p},\bw{x})
\bw{B}$ are the following terms
\beq
\begin{array}{l}\displaystyle
\left(\bw{\nabla}\bw{j}\right)_1=\nabla_i \left(\Omega^j
\frac{\partial
\stackrel{\tiny\downarrow}{n}({\bw{p},\bw{x}})}{\partial
p^j}\right)\epsilon({\bw{p},\bw{x}}) B^i,\\ \\ \displaystyle
\left(\bw{\nabla}\bw{j}\right)_2=\nabla_i \left(\Omega^j
\frac{\partial n({\bw{p},\bw{x}})}{\partial
p^j}\right)\stackrel{\tiny\downarrow}{\epsilon}({\bw{p},\bw{x}})
B^i,
\end{array}
\eeq
here the vertical arrows point out where the differential operator
acts. It is clear from elementary vector algebra identity
$\vec{k}(\vec{k}\times \vec{n})=0$ that the contribution of
$\nabla \left[ \epsilon({\bw{p},\bw{x}}) \left(\bw{\Omega}\times
\frac{\partial n({\bw{p},\bw{x}})}{\partial \bw{x}}\right)\right]$
is zero since the wave of density as well as the wave of energy
fluctuations propagate in the same direction of the wave-vector
$\vec{k}$.
The density waves will be sought in the same form as prescribed by
Landau theory for zero sound. We notice that such fluctuations,
assuming $\bw{k}\,||\, \bw{B}$, are precisely of the form that the
chiral spiral wave predicted from holography
\beq
\begin{array}{l} \displaystyle
\delta n_{R,L}(\theta,\phi) = \delta(\epsilon-\epsilon_F)
\nu_{R,L}(\theta,\phi) e^{i(\omega t - \bw{k}\bw{r})}.\\
\displaystyle
\end{array}
\eeq
For simplicity we stay in the axially symmetric sector
$\nu_{R,L}=\nu_{R,L}(\theta)$; certainly there will exist other
zero sound modes corresponding to decomposition in higher
harmonics according to the azimuthal angle $\phi$. Since we deal
with a system ``left fermions + right fermions + other degrees of
freedom'', we cannot make a precise statement on how the
fluctuations of the densities correspond to the energy
fluctuations.  Thus we model it as
\beq
\begin{array}{l}\displaystyle
\delta \epsilon_R=\int \left(F_S \nu_R(\theta')+F_A
\nu_L(\theta')\right)\frac{\sin\theta' d\theta'}{2},\\
\displaystyle \delta \epsilon_L=\int \left(F_A \nu_R(\theta')+F_S
\nu_L(\theta')\right)\frac{\sin\theta' d\theta'}{2}
\end{array}\eeq
The ``interaction form factors'' $F_S,F_A$ contain essentially all
the interesting dynamics of the problem. Here we chose the
simplest case with $F_{S,A}=const$, corresponding to the lowest
Legendre polynomials in the spherical functions expansion.
Defining them dimensionless as above, we allow ourselves to speak
about a ``weakly coupled system'' where $F_{S,A}\to 0$, and a
``strongly coupled system'' where $F_{S,A}\to\infty$. Anticipating
the next section of this work we should warn the reader that it is
not at all evident that the ``strongly coupled'' Landau model of
the Fermi liquid based on the nonrelativistic Fermi sphere picture
and the collisionless kinetic equation is not {\it a priori}
deemed to be equivalent to the holographic model of the same
object.

Under the definitions given above, the two contributions in
questions become \beq
\begin{array}{l}\displaystyle
\left(\bw{\nabla}\bw{j}_R(\theta)\right)_1=\frac{ik
\nu_R(\theta)\delta(\epsilon-\epsilon_F)  B v_F }{4p_F^2},\\
\\ \displaystyle \left(\bw{\nabla}\bw{j}_R(\theta)
\right)_2=\frac{ik \delta(\epsilon-\epsilon_F) B v_F }{2p_F^2}\int
\left(F_S\nu_R(\theta')+F_A\nu_L(\theta')\right) \frac{\sin\theta'
d\theta'}{2}.
\end{array}
\eeq
and analogously we can write down the contributions for the left
current $\bw{j}_L$. In the equation for
$\left(\bw{\nabla}\bw{j}\right)_1$ we have used an integration by
parts trick in order to avoid nasty terms like
$\nu'_{R,L}(\theta)$. Thus the equation\myref{zerosound} is
transformed into
\beq
\begin{array}{l}\displaystyle
(s-\cos\theta - b)\nu_R(\theta)=(b+\cos\theta)(F_S C_R+F_A C_L),\\
\displaystyle (s-\cos\theta - b)\nu_L(\theta)=(b+\cos\theta)(F_A
C_R+F_S C_L),
\end{array} \eeq
where a convenient dimensionless parameter
\beq b\equiv\frac{B v_F}{2p_F^2} \eeq
has been introduced. Solving this system with regard to
$\nu_{R,L}$ and imposing the conditions
\beq\begin{array}{l} \displaystyle \int\frac{\sin\theta^\prime
d\theta^\prime}{2}\nu_R(\theta^\prime;C_R,C_L)=C_R\\ \displaystyle
\int \frac{\sin\theta^\prime
d\theta^\prime}{2}\nu_L(\theta^\prime; C_R,C_L)=C_L
\end{array}
\eeq
we get thus a homogeneous system upon the coefficients $C_{R,L}$.
Requiring its determinant to be zero we obtain the dispersion
laws. They can conveniently be written down for the case of small
and large coupling. For small coupling we see that the degeneracy
has indeed been lifted
\beq s=1\pm b + e^{-\frac{2}{F_S(1\mp b)}}.\eeq
At zero magnetic field the Landau result is restored (obtainable
from\myref{disp1} taking $F_1=0, F_S=F_A=F_0$). If we look at the
eigenfunctions it can be shown that the axial zero sound is
proportional to the magnetic field while the vector zero sound
acquires the correction to the usual velocity.

At large coupling we get
\beq s=\sqrt{\frac{F_S}{3}\left(1-\frac{F_A^2-F_S^2}{2F_S}\right)}
\eeq
Again, we comply with the Landau result\myref{disp2} taking
$F_0=F_S\gg 1,F_S=F_A$. Notice that in both cases the field $B$
has been held arbitrary; the only perturbative expansion used was
the expansion in small or large $F_{S,A}$. We did not consider
here the non-anomalous effects of the magnetic field on the
structure of the modes, since in the linear order in the magnetic
field the non-anomalous effect will be upon spin waves rather than
zero sound waves.

These dispersion laws are the main result of the field-theoretical
part of this work. We interpret this situation in terms of two
types of zero sound, axial and vector one propagating through a
medium of left and right fermions with zero net chirality, which
interact due to the presence of the $B$ field. Below we analyze
the two zero sounds (or the anomalous zero sound) from the
holographic point of view. As we have remarked above the same
is true for the rotating case.

\section{Anomalous zero sound in holography}
\subsection*{Anomalous splitting}
Recently a great deal of activity has taken place, advocating
understanding quantum liquids (both of Fermi and non-Fermi type)
from holography; for some of the  results see
\mycite{Hartnoll:2008vx,Balasubramanian:2008dm,Cubrovic:2009ye,
Faulkner:2009wj,Ammon:2011hz,Faulkner:2010tq,Hartnoll:2009ns}; a
full review of the AdS/condensed state correspondence would
certainly go beyond the scope of the present paper; for a
pedagogic introduction see e.g.\mycite{Hartnoll:2009sz}.

\begin{figure}[h!]\begin{center}
\includegraphics[height = 5cm, width=5cm]{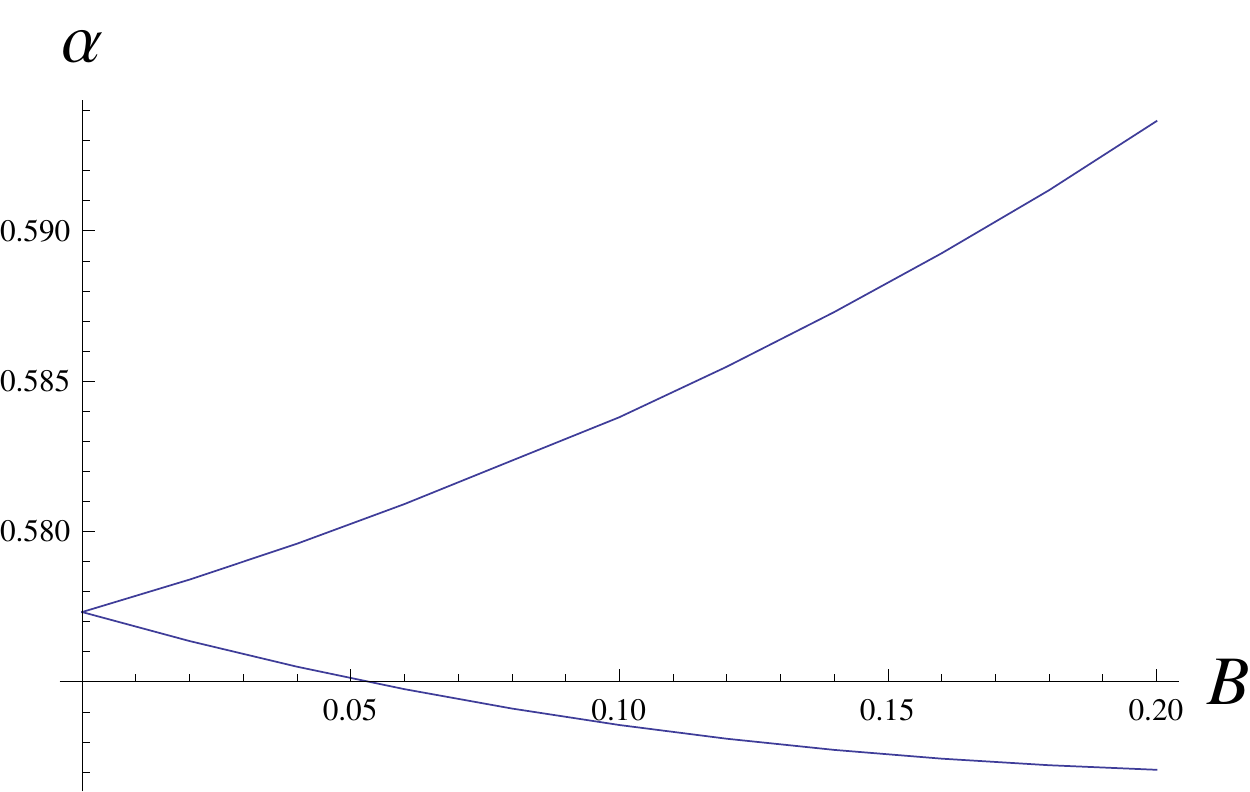}
\caption{\label{h1} Wave velocity $\alpha(B)$ of both modes of the
anomalous zero sound as a function of the magnetic field $B$.
}\end{center}
\end{figure}

\begin{figure}[h!]\begin{center}
\includegraphics[height = 5cm, width=5cm]{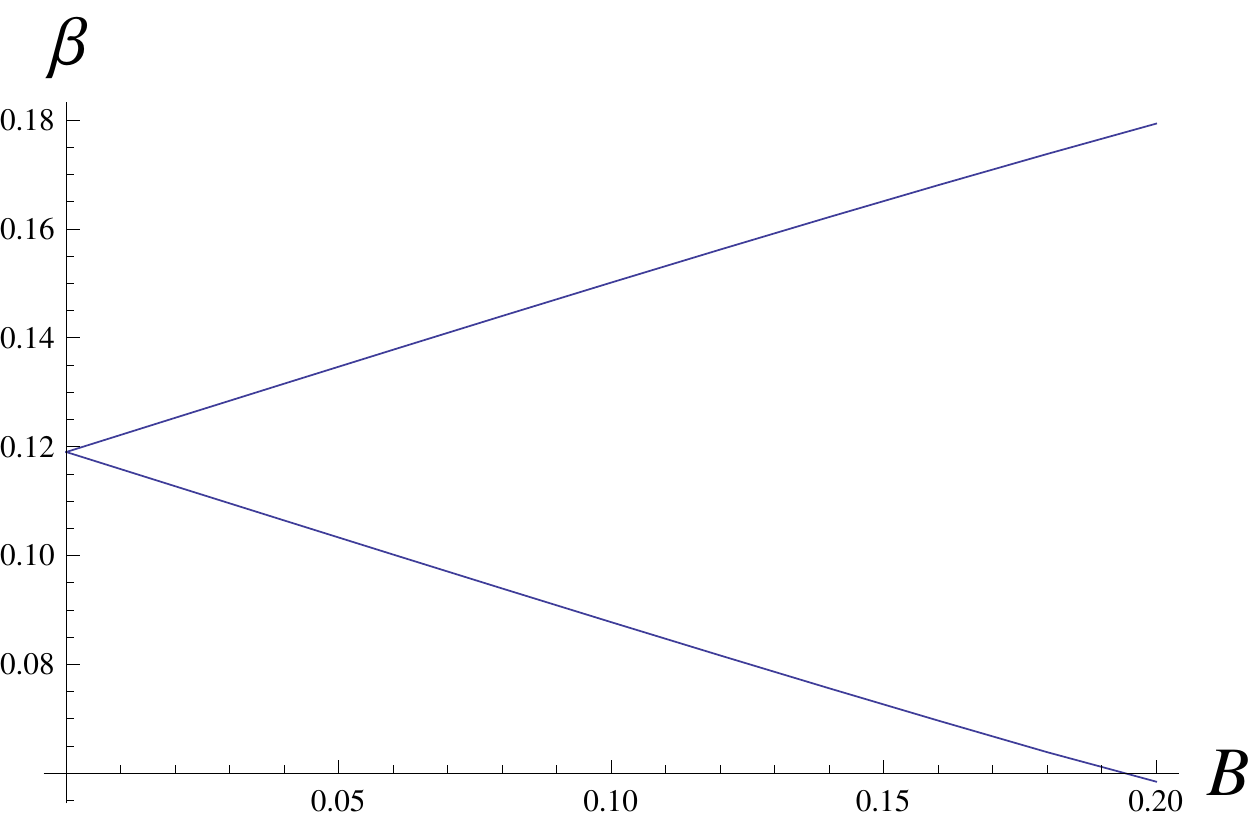}
\caption{\label{h2}Diffusion coefficient  $\beta(B)$ of both modes
of the anomalous zero sound as a function of the magnetic field
$B$. }\end{center}
\end{figure}

One can try to elucidate the nature of the mode we have observed
from the holographic point of view. The normal zero sound (i.e.
without the anomaly) in holography has been discovered by Karch,
Son and Starinets\mycite{Karch:2008fa}. It can be easily modified
to account for the anomalous effects, resident in the Chern-Simons
term. Consider a holographic model with the action
\beq S=S^L_{DBI}+S_{DBI}^{R}+S^L_{CS}-S_{CS}^{R} \eeq
where the (Dirac--Born--Infeld-motivated) action for either left
or right modes is
\beq S_{DBI}^{L,R}=-T\int d^8\xi
\sqrt{-\mathrm{det}\left(g_{ij}+2\pi \alpha' F_{ij}^{L,R}\right)}
\eeq
which in the metric
\beq ds^2=\frac{r^2}{R^2} dx_{1,3}^2+
\frac{R^2}{r^2}\left(dr^2+r^2 d\Omega_5^2\right)\eeq
in presence of the vector-potentials $A_0^{L,R}(r)$, and a
constant field $F_{12}^R=F_{12}^L=B$ becomes
\beq S=-\mathcal{N} V_0 \int dr \sqrt{\left(1-\left(\frac{\partial
A_0^{L,R}}{\partial r}\right)^2\right)(r^4+B^2)} \eeq
(here we have included the $2\pi\alpha'$ into the definitions of
the fields).
The Chern--Simons part is
\beq S_{CS}=-\frac{N_c}{24 \pi^2}\int dr d^4x
\epsilon^{\mu\nu\lambda\rho\sigma} A_\mu F_{\nu\lambda}
F_{\rho\sigma} .\eeq
Let us consider a system analogous to our model in the
four-dimensional part of this work. We have already mentioned
two types of holographically represented matter; now we must say
something about the chemical potential. To comply with the
previous sections we choose $\mu_L=\mu_R\equiv \mu$. Thus we can
solve the classical equations of motion for the $L$ and $R$ modes
independently by the Ansatz
\beq A^{L,R\prime}_{0,cl}(r)=\frac{d}{\sqrt{d^2+r^6+r^2B^2}} \eeq
where the integration constant $d\sim \mu^{\frac{1}{3}}$. When
fluctuations
\beq
\begin{array}{l}\displaystyle
A^{L,R}_{0}(r)=A^{L,R}_{0,cl}(r)+a_0^{L,R}(r,x,t)\\ \displaystyle
A^{L,R}_{3}(r)=a_3^{L,R}(r,x,t)\end{array}\eeq
are taken into account, the Chern--Simons term would not have
contributed to the fluctuation equations had three been a single
type of fermions solely. Yet due to having both left and right
modes, we can extract the $VVA$ ``anomalous  vertex'' structure
and obtain the following modification of the equations of motion
(25)a-c in\mycite{Karch:2008fa}
\beq
\begin{array}{l} \displaystyle
\partial_z\left(f^3 g a_0^{V\prime}\right)-\frac{qf}{z}
(\omega a_3^V+q a_0^V)+i b a_3^{A\prime}=0,\\ \displaystyle
\partial_z\left(f g a_3^{V\prime}\right)+\frac{\omega f}{z}
(\omega a_3^V+q a_0^V)-i b a_0^{A\prime}=0,\\ \displaystyle
\partial_z\left(f^3 g a_0^{L\prime}\right)-\frac{qf}{z}
(\omega a_3^A+q a_0^A)-i b a_3^{V\prime}=0,\\ \displaystyle
\partial_z\left(f g a_3^{A\prime}\right)+\frac{\omega f}{z}
(\omega a_3^A+q a_0^A)+i b a_0^{V\prime}=0,\\
\end{array}
\eeq
where
\beq a_{0,3}^{L,R}(r,x)=\int \frac{dq d\omega
}{(2\pi)^2}a_{0,3}^{L,R}(r,\omega,q )e^{-i\omega t + i q x},\eeq
the remnant gauge fixing condition being
\beq f^2\omega a_0^{V,A \prime}+q a_3^{V,A \prime}=0,\eeq
where we have switched to the variable $z\equiv\frac{1}{r}$ and
introduced the functions
\beq f=\sqrt{1+\frac{d^2z^6}{1+z^4B^2}},\eeq
and
\beq g=\frac{\sqrt{1+B^2 z^4}}{z}\eeq
and absorbed the normalization factor into $b\sim B$. Then
following\mycite{Karch:2008fa} we introduce two gauge-invariant
field strengths
\beq
\begin{array}{l}
E^V=\omega a_3^V+q a_0^V\\
E^A=\omega a_3^A+q a_0^A
\end{array}
\eeq
and construct a complex variable out of them
\beq E=E^V+i E^A ,\eeq
now representing the dynamics in the chiral plane, thus we can
finally write down the equation upon the complex variable  $E$
\beq E''(z)+E'(z) \left(\frac{f'(z) \left(3 q^2-\omega ^2
f(z)^2\right)}{f(z) \left(q^2-\omega^2
f(z)^2\right)}+\frac{g^\prime(z)}{g(z)}\right)+\frac{E(z)
\left(\omega^2 f(z)^2-q^2\right)}{f(z)^2}+\frac{i b q \omega z
E'(z)}{f(z) \left(q^2-\omega^2 f(z)^2\right)}=0 \eeq
This equation is then studied numerically. It is subject to the
boundary equations
\beq\left\{\begin{array}{l}
E(z)|_{z\to 0}=0\\
\partial_z (z E(z))|_{z\to\infty}=i\omega z E(z)|_{z\to\infty}
\end{array}\right.\eeq
The first of these conditions is the normal Dirichlet boundary
condition on the boundary; the second one is the infalling wave
condition on the horizon. These two conditions fix eigenvalues of
this equation
\beq \omega=\omega(k)=\alpha(B) k -i \beta (B) k^2.\eeq
We chose the Ansatz for $\omega(k)$ containing a linear and a
quadratic term only, which is confirmed by numeric calculations.
It can be also verified by numerical analysis that the term linear
in $k$ is indeed purely real whereas the term quadratic in $k$ is
purely imaginary. The numerical dependencies of the phase velocity
of the wave $\alpha$ and the diffusion coefficient $\beta$ on the
magnetic field are given in\myfigref{h1} and \myfigref{h2}
respectively for both modes of the anomalous zero sound.
One can readily see that the $\alpha$ dependence starts with a
correct $1/\sqrt{3}$ value, and then changes linearly for small
values of the field.  This is surely the Chern-Simons
contributions, since the DBI could have contributed only in the
order of $\sim B^2$. The field lifts the degeneracy, and the
splitting between the modes for small values of the field is
linear for small fields; at large fields the DBI terms quadratic
in $B$ start to contribute.

Thus we have produced in this section the dispersion relation for
the chiral anomalous zero sound in holography generalizing
analysis of \mycite{Karch:2008fa}. Notice that the fluctuation
effect that we have described exists independently from the static
anomalous axial current that will be present in our setting as
well\mycite{Rebhan:2010ax} but has not been analyzed here as in
the leading order the two effects are essentially decoupled.

\section{Discussion}
In this note we have discussed the modification of the zero sound
excitation of the Fermi surface due to the anomaly in the external
magnetic field. Since the vector and axial modes are coupled in
the magnetic field we get two independent zero sound modes instead
of the single one. Using the nontrivial Berry phase in the
momentum space we get the dispersion relations for both zero sound
modes in the framework of the kinetic equations.

Our study was focused on the peculiar longitudinal modes and a
detailed analysis for the transverse modes can be taken into
account as well. We have considered only the fluctuations of the
components of the vector and axial currents ignoring the spin
effects due to the fluctuations of the tensor currents. There
should be interesting effect on the zero spin sound due to the
mixing of scalar and tensor modes in the external magnetic field
discussed in\mycite{Gorsky:2012ui}. Another study case for us
would be the involvement of the spin-spin interaction and
influence of the magnetic field  on interaction between the spin
waves and sound waves. It would be also useful to discuss
separately the effects of the possible Fermi points in the spin
sound case.

Our consideration has a lot in common with the chiral spiral
wave\mycite{Kharzeev:2011rw} however in that case the modes in
plasma have been investigated implying the high temperature while
in our case we discuss the dense matter at very small or vanishing
temperature. At the holographic side we have generalized the
consideration of\mycite{Goykhman:2012vy} taking into account the
Chern-Simons term and have found the clear-cut manifestation of
the anomalous contributions at the strong coupling. It would be
interesting to investigate the possibility of the experimental
observation of the magnetic or rotational anomalous zero sound.

\section*{Acknowledgements}
We would like to thank A. Parnachev, G. Policastro and A.
Starinets for useful discussions. Special thanks to Niko Jokela
and Ho-Ung Yee for their valuable comments. The work of A.Z. is
supported in part by the RFBR grants 10-01-00836 and 10-02-01483
supported by Ministry of Education and Science of the Russian
Federation under the contract 14.740.11.0081. A.Z. thanks the
organizers and the participants of the GGI-Florence 2011 workshop
on Large-N Gauge Theories where some of the ideas behind the
present work originated. He also thanks D.Kharzeev for warm
hospitality at Stony Brook and stimulating discussions. The work
of A.G. is supported in part by grants RFBR-12-02-00284 and PICS-
12-02-91052


\begin{thebibliography}{10}

\bibitem{Landau1957}
L.~Landau, ``{The theory of a Fermi liquid},'' {\em Zh. Eksp.
Teor. Fiz.} {\bf
  30} (1956)  1058.

\bibitem{Karch:2008fa}
A.~Karch, D.~Son, and A.~Starinets, ``{Zero Sound from
Holography},'' \href{http://arxiv.org/abs/0806.3796}{{\tt
arXiv:0806.3796 [hep-th]}}.

\bibitem{Son:2012wh}
D.~T. Son and N.~Yamamoto, ``{Berry Curvature, Triangle Anomalies,
and Chiral
  Magnetic Effect in Fermi Liquids},''
\href{http://arxiv.org/abs/1203.2697}{{\tt arXiv:1203.2697
  [cond-mat.mes-hall]}}.

\bibitem{Fukushima:2008xe}
K.~Fukushima, D.~E. Kharzeev, and H.~J. Warringa, ``{The Chiral
Magnetic
  Effect},'' \href{http://dx.doi.org/10.1103/PhysRevD.78.074033}{{\em
  Phys.Rev.} {\bf D78} (2008)  074033},
\href{http://arxiv.org/abs/0808.3382}{{\tt arXiv:0808.3382
[hep-ph]}}.

\bibitem{Yee:2009vw}
H.-U. Yee, ``{Holographic Chiral Magnetic Conductivity},''
  \href{http://dx.doi.org/10.1088/1126-6708/2009/11/085}{{\em JHEP} {\bf 0911}
  (2009)  085},
\href{http://arxiv.org/abs/0908.4189}{{\tt arXiv:0908.4189
[hep-th]}}.

\bibitem{Rebhan:2009vc}
A.~Rebhan, A.~Schmitt, and S.~A. Stricker, ``{Anomalies and the
chiral magnetic
  effect in the Sakai-Sugimoto model},''
  \href{http://dx.doi.org/10.1007/JHEP01(2010)026}{{\em JHEP} {\bf 1001} (2010)
   026},
\href{http://arxiv.org/abs/0909.4782}{{\tt arXiv:0909.4782
[hep-th]}}.

\bibitem{Gorsky:2010xu}
A.~Gorsky, P.~Kopnin, and A.~Zayakin, ``{On the Chiral Magnetic
Effect in
  Soft-Wall AdS/QCD},''
  \href{http://dx.doi.org/10.1103/PhysRevD.83.014023}{{\em Phys.Rev.} {\bf D83}
  (2011)  014023},
\href{http://arxiv.org/abs/1003.2293}{{\tt arXiv:1003.2293
[hep-ph]}}.

\bibitem{Buividovich:2009wi}
P.~Buividovich, M.~Chernodub, E.~Luschevskaya, and M.~Polikarpov,
``{Numerical
  evidence of chiral magnetic effect in lattice gauge theory},''
  \href{http://dx.doi.org/10.1103/PhysRevD.80.054503}{{\em Phys.Rev.} {\bf D80}
  (2009)  054503},
\href{http://arxiv.org/abs/0907.0494}{{\tt arXiv:0907.0494
[hep-lat]}}.

\bibitem{Kharzeev:2010gd}
D.~E. Kharzeev and H.-U. Yee, ``{Chiral Magnetic Wave},''
  \href{http://dx.doi.org/10.1103/PhysRevD.83.085007}{{\em Phys.Rev.} {\bf D83}
  (2011)  085007},
\href{http://arxiv.org/abs/1012.6026}{{\tt arXiv:1012.6026
[hep-th]}}.

\bibitem{Kirilin:2012sd}
V.~Kirilin, Z.~Khaidukov, and A.~Sadofyev, ``{Chiral Vortical
Effect in Fermi
  Liquid},''
\href{http://arxiv.org/abs/1203.6612}{{\tt arXiv:1203.6612
  [cond-mat.mes-hall]}}.

\bibitem{Zahed:2012yu}
I.~Zahed, ``{Anomalous Chiral Fermi Surface},''
\href{http://arxiv.org/abs/1204.1955}{{\tt arXiv:1204.1955
[hep-th]}}.

\bibitem{Baym:1975va}
G.~Baym and S.~A. Chin, ``{Landau Theory of Relativistic Fermi
Liquids},''
\href{http://dx.doi.org/10.1016/0375-9474(76)90513-3}{{\em
Nucl.Phys.} {\bf
  A262} (1976)  527}.

\bibitem{Hartnoll:2008vx}
S.~A. Hartnoll, C.~P. Herzog, and G.~T. Horowitz, ``{Building a
Holographic
  Superconductor},''
  \href{http://dx.doi.org/10.1103/PhysRevLett.101.031601}{{\em Phys.Rev.Lett.}
  {\bf 101} (2008)  031601},
\href{http://arxiv.org/abs/0803.3295}{{\tt arXiv:0803.3295
[hep-th]}}.

\bibitem{Balasubramanian:2008dm}
K.~Balasubramanian and J.~McGreevy, ``{Gravity duals for
non-relativistic
  CFTs},'' \href{http://dx.doi.org/10.1103/PhysRevLett.101.061601}{{\em
  Phys.Rev.Lett.} {\bf 101} (2008)  061601},
\href{http://arxiv.org/abs/0804.4053}{{\tt arXiv:0804.4053
[hep-th]}}.

\bibitem{Cubrovic:2009ye}
M.~Cubrovic, J.~Zaanen, and K.~Schalm, ``{String Theory, Quantum
Phase
  Transitions and the Emergent Fermi-Liquid},''
  \href{http://dx.doi.org/10.1126/science.1174962}{{\em Science} {\bf 325}
  (2009)  439--444},
\href{http://arxiv.org/abs/0904.1993}{{\tt arXiv:0904.1993
[hep-th]}}.

\bibitem{Faulkner:2009wj}
T.~Faulkner, H.~Liu, J.~McGreevy, and D.~Vegh, ``{Emergent quantum
criticality,
  Fermi surfaces, and AdS(2)},''
  \href{http://dx.doi.org/10.1103/PhysRevD.83.125002}{{\em Phys.Rev.} {\bf D83}
  (2011)  125002},
\href{http://arxiv.org/abs/0907.2694}{{\tt arXiv:0907.2694
[hep-th]}}.

\bibitem{Ammon:2011hz}
M.~Ammon, J.~Erdmenger, S.~Lin, S.~Muller, A.~O'Bannon, {\em et
al.}, ``{On
  Stability and Transport of Cold Holographic Matter},''
  \href{http://dx.doi.org/10.1007/JHEP09(2011)030}{{\em JHEP} {\bf 1109} (2011)
   030},
\href{http://arxiv.org/abs/1108.1798}{{\tt arXiv:1108.1798
[hep-th]}}.

\bibitem{Faulkner:2010tq}
T.~Faulkner and J.~Polchinski, ``{Semi-Holographic Fermi
Liquids},''
  \href{http://dx.doi.org/10.1007/JHEP06(2011)012}{{\em JHEP} {\bf 1106} (2011)
   012},
\href{http://arxiv.org/abs/1001.5049}{{\tt arXiv:1001.5049
[hep-th]}}.

\bibitem{Hartnoll:2009ns}
S.~A. Hartnoll, J.~Polchinski, E.~Silverstein, and D.~Tong,
``{Towards strange
  metallic holography},'' \href{http://dx.doi.org/10.1007/JHEP04(2010)120}{{\em
  JHEP} {\bf 1004} (2010)  120}, \href{http://arxiv.org/abs/0912.1061}{{\tt
  arXiv:0912.1061 [hep-th]}}.
71 pages, 8 figures.

\bibitem{Hartnoll:2009sz}
S.~A. Hartnoll, ``{Lectures on holographic methods for condensed
matter
  physics},'' \href{http://dx.doi.org/10.1088/0264-9381/26/22/224002}{{\em
  Class.Quant.Grav.} {\bf 26} (2009)  224002},
\href{http://arxiv.org/abs/0903.3246}{{\tt arXiv:0903.3246
[hep-th]}}.

\bibitem{Rebhan:2010ax}
A.~Rebhan, A.~Schmitt, and S.~Stricker, ``{Holographic chiral
currents in a
  magnetic field},'' \href{http://dx.doi.org/10.1143/PTPS.186.463}{{\em
  Prog.Theor.Phys.Suppl.} {\bf 186} (2010)  463--470},
\href{http://arxiv.org/abs/1007.2494}{{\tt arXiv:1007.2494
[hep-th]}}.

\bibitem{Jokela:2012vn}
N.~Jokela, G.~Lifschytz, and M.~Lippert, ``{Magnetic effects in a
holographic
  Fermi-like liquid},'' {\em JHEP} {\bf 1205} (2012)  105,
\href{http://arxiv.org/abs/1204.3914}{{\tt arXiv:1204.3914
[hep-th]}}.

\bibitem{Goykhman:2012vy}
M.~Goykhman, A.~Parnachev, and J.~Zaanen, ``{Fluctuations in
finite density
  holographic quantum liquids},''
\href{http://arxiv.org/abs/1204.6232}{{\tt arXiv:1204.6232
[hep-th]}}.

\bibitem{Gorsky:2012ui}
A.~Gorsky, P.~Kopnin, A.~Krikun, and A.~Vainshtein, ``{More on the
Tensor
  Response of the QCD Vacuum to an External Magnetic Field},'' {\em Phys.Rev.}
  {\bf D85} (2012)  086006,
\href{http://arxiv.org/abs/1201.2039}{{\tt arXiv:1201.2039
[hep-ph]}}.

\bibitem{Kharzeev:2011rw}
D.~E. Kharzeev and H.-U. Yee, ``{Chiral helix in AdS/CFT with
flavor},''
  \href{http://dx.doi.org/10.1103/PhysRevD.84.125011}{{\em Phys.Rev.} {\bf D84}
  (2011)  125011},
\href{http://arxiv.org/abs/1109.0533}{{\tt arXiv:1109.0533
[hep-th]}}.

\end{thebibliography}

\providecommand{\href}[2]{#2}\begingroup\raggedright\endgroup

\end{document}